\documentclass[apjl]{emulateapj}         

\usepackage{graphics}

\newcommand{\um}{${\rm \mu m}$~}
\newcommand{\mm}{${\rm \mu m}$}

\long\def\symbolfootnote[#1]#2{\begingroup%
\def\thefootnote{\fnsymbol{footnote}}\footnote[#1]{#2}\endgroup}

\slugcomment{accepted for publication in ApJ} 

\begin{document}

\title{Debris Distribution in HD 95086 -- A Young Analog of HR 8799}
 \author{Kate Y. L. Su\altaffilmark{1},
   Sarah Morrison\altaffilmark{2},
   Renu Malhotra\altaffilmark{2},
   Paul S. Smith\altaffilmark{1}, 
   Zoltan Balog\altaffilmark{3},
   George H. Rieke\altaffilmark{1}
  }

\altaffiltext{1}{Steward Observatory, University of Arizona, 933 N
  Cherry Ave., Tucson, AZ 85721; ksu@as.arizona.edu}
\altaffiltext{2}{Lunar and Planetary Laboratory, University of
  Arizona, Tucson, AZ 85721, USA}
\altaffiltext{3}{Max-Planck-Institut f\"ur Astronomie, K\"onigstuhl 17
  D-69117, Heidelberg, Germany}

\begin{abstract}

HD 95086 is a young early-type star that hosts (1) a 5 $M_J$ planet at
the projected distance of 56 AU revealed by direct imaging, and (2) a
prominent debris disk. Here we report the detection of 69 \um
crystalline olivine feature from the disk using the {\it
Spitzer}/MIPS-SED data covering 55--95 \mm. Due to the low resolution
of MIPS-SED mode, this feature is not spectrally resolved, but is
consistent with the emission from crystalline forsterite contributing
$\sim$5\% of the total dust mass. We also present detailed analysis of
the disk SED and re-analysis of resolved images obtained by {\it
Herschel}. Our results suggest that the debris structure around HD
95086 consists of a warm ($\sim$175 K) belt, a cold ($\sim$55 K) disk,
and an extended disk halo (up to $\sim$800 AU), and is very similar to
that of HR 8799. We compare the properties of the three debris
components, and suggest that HD 95086 is a young analog of HR 8799. We
further investigate and constrain single-planet, two-planet,
three-planet and four-planet architectures that can account for the
observed debris structure and are compatible with dynamical stability
constraints.  We find that equal-mass four-planet configurations of
geometrically spaced orbits, with each planet of mass $\sim$5 $M_J$, could
maintain the gap between the warm and cold debris belts, and also be
just marginally stable for timescales comparable to the age of the
system.

\end{abstract} 

               
\section{Introduction}

Hundreds of planetary systems have been discovered through radial
velocity \citep{mayor11} and transit measurements
\citep{burke14}. However, this breakthrough is currently biased toward
planetary architectures that may have experienced much more dynamical
evolution and reflect conditions that are different from our own Solar
System. Probing planets similar to our own Solar System (giant planets
located outside 5 AU with orbital periods longer than 10 years)
requires either long-term monitoring or advanced instruments that can
significantly reduce the light from the host stars, which are both
technically challenging. Debris disks offer an indirect way to study
outer planetary systems. Debris disks are tenuous dusty structures
sustained through collisions of leftover planetesimals and minor
bodies such as asteroids, Kuiper Belt Objects (KBOs) and comets (often
referred as parent bodies), that failed to form planets. The loss
timescale for dust in a typical debris system is generally less than
10$^4$ years; therefore, the presence of dust around a main-sequence
star requires a replenishing reservoir and one or more large bodies
providing dynamical stirring to maintain a high rate of collisions
among the debris parent-body reservoir.  The large surface area of
debris makes these disks detectable through infrared and millimeter
thermal emission or optical scattered light, providing insights into
the nature of unseen parent-body populations and massive planetary
perturbers. Because debris disks are detected from just after the
protoplanetary stage to nearly the end of the star's life billions of
years later, they are excellent observational tools to study the
growth of planets and subsequent dramatic steps that determine the
architecture of planetary systems.

In our Solar System, the stable locations of the leftover
planetesimals (i.e., the asteroid and Kuiper belts) are tightly
coupled with the architecture of the planets. In an analogous fashion,
the observed debris structures around other stars such as the tilted
inner disk in $\beta$ Pic \citep{heap00,golimowski06} and the
eccentric ring in Fomalhaut \citep{stapelfeldt04,kalas05} have long
been used to predict the presence of unseen planets.  This connection
became even more persuasive when the first few directly imaged planets
were discovered around debris-host stars (HR 8799: \citealt{marois08},
Fomalhaut: \citealt{kalas08}, and $\beta$ Pic:
\citealt{lagrange09}). In addition, some debris systems are known to
possess dust emission at two different temperatures, suggesting a cold
outer belt accompanied by a fainter, warm inner one (e.g.,
\citealt{chen09,morales11,ballering13,kennedy14}).  Although there
might be many mechanisms to explain such a large gap, a probable way
to maintain this structure is shown by the four super-Jupiter-like
planets around HR 8799 \citep{marois10}, which are packed between
inner and outer debris disk components as revealed by resolved imaging
and detailed analysis of the spectral energy distribution (SED)
\citep{su09}. The similar large gaps between the inner and outer
debris components around other stars may be maintained by lower-mass
planets; for example, in the debris disk twins (Vega and Fomalhaut),
we have strong evidence that the large gap between the warm
asteroid-like and the cold KBO-like dust belts is an excellent
signpost for multiple ice-giant-mass planets beyond the ice line
\citep{su13}. A large gap of this type between the inner warm and
outer cold belts is an example of a global disk feature that points
toward general aspects of planetary system formation and evolution.

The SEDs of debris disks can be understood, to first order, by
assuming that there are various zones that are maintained by common
factors, e.g., fossil ice lines and ice giant planets. The diversity
of disk structures and SEDs then reflect variations in the amount of
material in these zones, that are closely related to different paths
by which a system forms and evolves, rather than radical differences
in structure \citep{su14}. Hence, systems that harbor directly-imaged
planets and debris disks are valuable tools to better understand
planetary system formation and evolution.

In this study, we focus on the planetary system around HD 95086. The
star is an early-type member of the Lower Centaurus Crux Association
(LCC; \citealt{dezeeuw99,rizzuto12}), and located 90.4$\pm$3.4 pc away
\citep{vanleeuwen07}. The age of the star has been reviewed by
\citet{meshkat13} with an estimate of 17 Myr ($\pm$2 Myr statistical
and $\pm$4 Myr systematic). HD 95086 has drawn a lot of attention
lately because it hosts a directly imaged planet HD 95086b at a
projected distance of 56 AU \citep{rameau13a}. The planet has very red
infrared colors, similar to young massive planets HR 8799 bcde and 2M
1207b, with a mass estimate of 5$\pm$2 $M_J$ based on evolutionary
models \citep{meshkat13,rameau13b,galicher14}. HD 95086 also has a
prominent infrared excess with a dust fractional luminosity ($f_d$) of
1.6$\times10^{-3}$ \citep{rhee07,chen12}. The debris around HD 95086
was marginally resolved at far-infrared wavelengths with {\it
Herschel}, suggesting a slightly inclined orientation from face-on
\citep{moor13}. There are some inconsistencies in the disk properties
of HD 95086 in the literature. We, therefore, review existing infrared
photometric and spectroscopic data, and present an unpublished {\it
Spitzer} MIPS-SED low-resolution spectrum covering 55--95 \um for a
comprehensive disk SED analysis (Section 2). To thoroughly exploit the
opportunity that the HD 95086 system provides, we also re-analyze the
{\it Herschel} resolved images in comparison with the SED modeling
results, and conclude that the resolved debris structure is mostly in
the form of a disk halo composed of small grains (Section 3). In
Section 4, we discuss the similarity between HD 95086 and HR 8799 in
their debris distribution, and provide dynamical constraints on
possible planetary configurations in HD 95086. Our conclusions are
presented in Section 5.

\section{Observations and Data Reduction} 
\label{obs} 

We present unpublished {\it Spitzer} MIPS-SED data and review all
published infrared/submillimeter observations for the HD 95086
system. Our goal is to construct a detailed disk SED that can be used
to estimate the location of debris within the system.

\subsection{Infrared/Submillimeter Broad-Band Photometry} 

{\it Spitzer} MIPS photometry was first published by \citet{chen12}
where fluxes were extracted using Point Spread Function (PSF) fitting
designed for point sources. We examined the 24 \um image, and found
the source is slightly extended (Full-Width-Half-Maximum (FWHM) of
5\farcs98$\times$5\farcs87, compared to a typical point source's FWHM
5\farcs49$\pm$0\farcs03$\times$5\farcs45$\pm$0\farcs03). Therefore, we
adopted measurements with photometry through a large aperture
(aperture radius of 14\farcs9 and sky annulus of
29\farcs9--42\farcs3), resulting in 52.95$\pm$0.55 mJy as the total 24
\um flux for the system ($\sim$16\% higher than the PSF extracted
flux). At 70 \mm, the source is point-like with a total flux of
655$\pm$33 mJy. The quoted flux uncertainties include instrumental
repeatability, which is 1\% and 5\% at 24 and 70 \mm, respectively.

{\it Herschel} PACS and SPIRE observations were first published by
\citet{moor13}, and showed that the disk was resolved at 70 and 100
\um with an estimated inclination angle of 25\arcdeg. Because the {\it
Herschel} observations were reduced with an older pipeline (HIPE
v9.2), we re-reduced the data with an updated pipeline (HIPE v13.0)
that incorporated many new calibration improvements detailed in
\citet{balog14}. Specifically, we included the experimental tool for
pointing reconstruction based on the gyroscope information that
considerably reduces the pointing jitter and increases the stability
of the PSF (termed as gyro correction). Finally, we mosaicked the
final maps with a default ``pixfrac'' parameter 1.0, equivalent to
drizzle parameter of 0.8, in the instrumental orientation so the PSF
structures were always in the same location. No rotation/interpolation
is then needed in further PSF subtraction analyses. Detailed analysis
of the PACS images is presented in Section 3.1. We used an aperture
size of 22\arcsec\ at all three bands to measure the source flux, at
which radius the encircled flux reached the maximum and flattened
afterward (after proper aperture correction). We did not apply color
correction for the PACS fluxes. The final PACS fluxes are: 688$\pm$48,
688$\pm$48, and 489$\pm$35 mJy at 70, 100, and 160 \mm, respectively,
in good agreement with the \citet{moor13} published results.

Inspecting the SPIRE maps, the source is clearly detected at all three
wavelengths as a point source, and the surrounding field has no bright
source or large-scale extended structure. Therefore, we used the level
2 point-source product (Jy/beam) provided by the Herschel Science
Center (HIPE ver.\ 11) to measure fluxes. We used the peak value of
the source to estimate the point-source flux. After correcting the
pixelisation effect (0.951, 0.931, 0.902 at 250, 350, and 500 \mm,
respectively; SPIRE Handbook, version 2.5, 2014), the source
submillimeter fluxes are: 186$\pm$16 mJy, 112$\pm$16 mJy, and
67$\pm$17 mJy at 250, 350, and 500 \mm, respectively.  At 500 \mm,
there are a couple of faint peaks within 40\arcsec\ of the target,
which might contaminate the submillimeter fluxes. No color correction
was applied for the SPIRE fluxes either. Our SPIRE fluxes are lower
than the values published by \citet{moor13} who extracted source
fluxes using aperture photometry, but consistent within
uncertainties. Finally, HD 95086 was detected at 870 \um with the
submillimeter telescope APEX by \citet{nilsson10}. Its nominal beam
size is 19\farcs2 (FWHM), but the effective resolution is 27\arcsec\
after filtering and smoothing in the data reduction
\citep{nilsson10}. The source at 870 \um appears to be elongated
(Figure 2 in \citealt{nilsson10}) and surrounded by faint
peaks. Therefore, it is likely that the measured 870 \um flux
(41.3$\pm$18.4 mJy) is over-estimated and contaminated by background
galaxies.

\begin{figure} 
  \figurenum{1}
  \label{exesed} 
  \epsscale{1.15} 
  \plotone{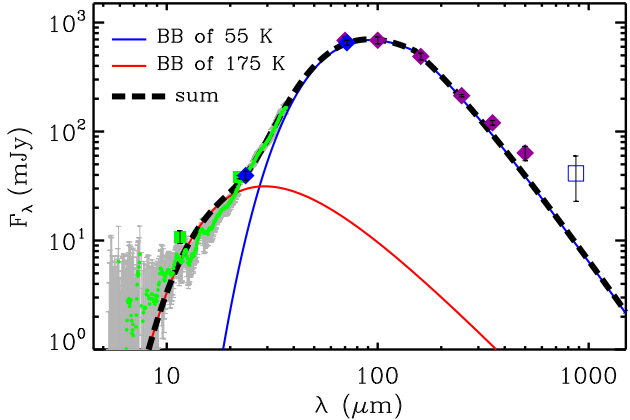} 
  \caption{The excess SED of the HD 95086 system composed of
broad-band photometry and the mid-infrared spectrum after subtraction
of the stellar contribution. The excess flux uncertainty (1$\sigma$)
includes 2\% of the photospheric emission added quadratically. The
broad-band fluxes include: blue diamonds from {\it Spitzer}, purple
diamonds from {\it Herschel}, green squares from {\it WISE} and a blue
open square from APEX/870 observation \citep{nilsson10}.}
\end{figure}

\subsection{Spectroscopy -- {\it Spitzer} IRS and MIPS-SED}

A {\it Spitzer} IRS spectrum of the system was first published by
\citet{chen06} where the excess spectrum was characterized by 
blackbody emission of 80$\pm$30 K. The same spectrum was reduced and
analyzed by \citet{moor13}, who found the spectral shape of the excess
along with long-wavelength photometry is better fitted with two
blackbody temperatures: 187$\pm$26 K and 57$\pm$1.5 K. Finally,
\citet{chen14} characterized the excess to be consistent with two
temperatures: 225$^{+10}_{-7}$ K and 57$\pm$5 K. The warm temperature
discrepancy might be due to how the two IRS modules were joined and scaled
relative to the photosphere. For this reason, we also performed the
IRS data reduction, similar to what was done in \citet{moor13} but
with the SMART software (ver.\ 8.2.7; \citealt{higdon04}). The two low
resolution modules (short-low (SL): 5.2--14.5 \um and long-low (LL):
14.0--38.0 \mm) were extracted and combined independently using
SMART's optimal 2 nod extraction \citep{lebouteiller10}. The SL
spectrum agrees very well ($<$1\%) with the expected Kurucz model flux
for wavelengths shorter than 6.5 \mm, while the LL spectrum was scaled
down (by 0.963) to match the MIPS 24 \um large aperture
photometry. The two modules then joined smoothly without further
scaling.

The photospheric contribution was subtracted using the best-fit Kurucz
model ($T_{eff}=7500 K$ and $log g=4.0$), and 2\% of the photospheric
emission was included in computing the excess flux uncertainty. Figure
\ref{exesed} shows the excess SED composed of broad-band photometry
and the IRS spectrum. Our best-fit (weighted by the uncertainty) two
blackbody temperatures are 55$\pm$5 K (cold component) and 175$\pm$25
K (warm component). The large error in the warm temperature is driven
by the large uncertainty in the 7--10 \um excess flux after
photospheric subtraction. The 175 K blackbody emission slightly
(within 1$\sigma$) under predicts the excess flux shortward of 10 \mm,
suggesting either the system has another faint, hotter component or
the emission from the warm component has a weak silicate feature from
$\sim$\mm-size grains. We have used the fitting technique developed by
\citet{ballering14}, including various normalizations of the
photosphere, to evaluate this emission. We find no evidence for
silicate emission features, thus differing from {\bf those} excesses
identified by \citet{ballering14}; instead, within the errors, a
$\sim$300 K blackbody is an adequate fit. Adding this component has
negligible effect on the fitted temperature of the 175 K component
(reducing its temperature by about 10 K). Given its large uncertainty,
we do not include this tentative, faint and hotter component in our
following discussion.

\begin{figure} 
  \figurenum{2}
  \label{exeMSED} 
  \epsscale{1.15} 
  \plotone{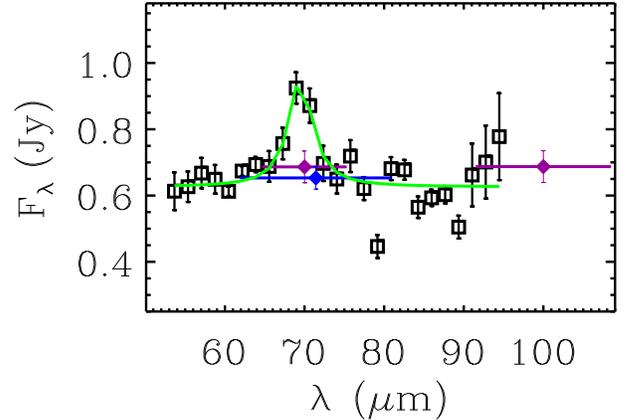} 
  \caption{The panel shows the excess MIPS-SED spectrum (black
squares). A crystalline olivine feature is detected where a Lorentzian
profile fit (green line) suggests the feature peaks at 69.5$\pm$0.5
\um and is not spectrally resolved. The MIPS-SED spectrum agrees well
with the broad-band photometry (diamonds). }
\end{figure}

\begin{figure*}
  \figurenum{3}
  \label{pacs_im} 
  \epsscale{1.15}
  \plotone{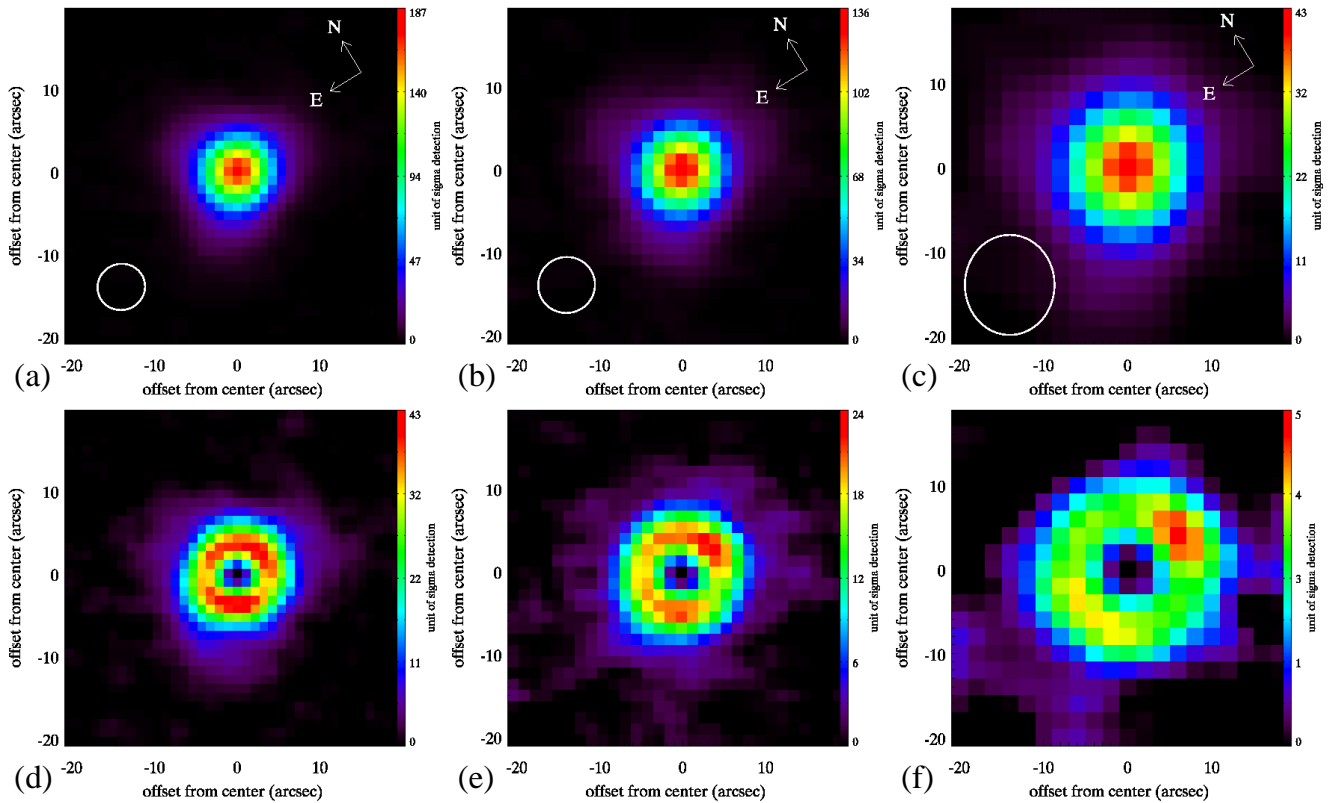}
  \caption{{\it Herschel} PACS images of HD 95086. The upper row shows
PACS images at (a) 70 \mm, (b) 100 \mm, and (c) 160 \mm,
respectively. The bottom row shows the PSF-subtracted (normalized to
the peak of the source) residual images at (d) 70 \mm, (e) 100 \mm,
and (f) 160 \mm, respectively. The beam sizes {\bf are} shown as white
ellipses in the upper panels.}
\end{figure*}

The MIPS-SED data were obtained in 2006 Feb 20 (AOR key 16171264, PID
84, PI: Jura) with the 10 s $\times$ 10 cycles and 1\arcmin\ chop
setting, resulting in a low-resolution ($R=$ 15--25) spectrum from 55
to 95 \um with 600 s of integration on source. The data were reduced
and calibrated as described by \citet{lu08}. An extraction aperture of
50\arcsec\ in the spatial direction was used, and the slit loss was
corrected assuming a point source. We trimmed off the spectrum past 95
\um because of known contamination by the blue end of the second-order
spectrum produced in this observing mode of the instrument that cannot
be calibrated (Lu et al. 2008). The photosphere-subtracted MIPS-SED
spectrum is shown in Figure \ref{exeMSED}, where a feature peaking
near $\sim$70 \um can be seen. There is also one low discrepant point
near 80 \um (by 2--3$\sigma$) that is problematic. We investigated the
discrepancy and were not able to find an obvious reason outside the
possible transient systematic errors in the array around 80
\mm. However, the feature at $\sim$70 \um appears to be real since it
consists of two high points that are $\gtrsim$3$\sigma$ above the
overall continuum, and several adjacent points that are consistent
with the instrumental resolution, unlike the single low point at 80
\mm. We have also checked other MIPS-SED spectra for sources that have
similar 70 \um fluxes as HD 95086, and none were found to have such an
emission feature, suggesting that it is not due to instrumental
artifacts. The shape of the MIPS-SED spectrum is consistent with the
broad-band photometry measured by PACS and MIPS at 70 \mm.  Since the
bandwidth of the MIPS 70 \um filter is wider than that of the PACS 70
\um filter, the slightly lower value of the MIPS 70 \um measurement is
consistent with the presence of the expected feature. A Lorentzian
profile fit to the MIPS-SED spectrum suggests that the feature peaks
at 69.5$\pm$0.5 \um with a linewidth (FWHM) of 4.2 \mm, consistent
with an emission-line feature that is not spectrally resolved in the
MIPS-SED mode. Given the cold dust temperature (55$\pm$5 K), the
measured peak position is consistent with the laboratory measurement
of crystalline olivine (Mg$_2$SiO$_4$, forsterite) at 50 K
\citep{suto06}. The 69 \um band strength (integrated luminosity) is
8.3$\times10^{-13}$ erg~cm$^{-2}$~s$^{-1}$, which is in the range of
detected band strengths around protoplanetary disks (see Table 2 in
\citealt{maaskant14}). This feature has also been detected in the
$\beta$ Pic debris disk using the {\it Herschel} PACS spectrometer
\citep{devries12}. We discuss the possible implications of the
presence of the 69 \um feature in Section 4.2.

\section{Analysis}

\subsection{PACS Imaging Analysis}

Fully understanding the instrumental PSFs is essential in
characterizing resolved images; especially it has been shown that
the high-pass filtering and drizzling methods for coadding images used in
HIPE have a great impact on the resultant PSF and noise
characteristics \citep{popesso12}. We, therefore, carried out some
basic PACS PSF characterization by using PACS data for $\gamma$ Dra, a
routine calibrator, that were observed throughout {\it Herschel}'s
operation. We used the data obtained in the mini-scan map mode
with two scan angles of 70\arcdeg\ and 110\arcdeg\ only (the same as
the HD 95086 data). These data were processed with the same pipeline
and procedure as our target star (same filtering width, masking
radius, pixfrac, instrumental orientation and gyro correction). We
further inspected the quality of the data by measuring the FWHM of
each observation and rejected data sets that have deviant
values. Finally, we obtained a high signal-to-noise, co-added $\gamma$
Dra image by median combining sets of 26, 5 and 26 observations at 70,
100 and 160 \mm, respectively. These combined, high-quality $\gamma$
Dra data were used as our point source standards when assessing the
resolvability of an observation.

All the FWHM measurements were measured by fitting a 2-D Gaussian
function on sub-regions centered at the target with fields of view
(f.o.v.) of 21\farcs1, 29\farcs4, and 44\farcs1 at 70, 100 and 160
\mm, respectively. Note that different values of f.o.v.\ would result
in slightly different FWHM values; therefore, it is important to be
consistent with the f.o.v. used in comparisons. The average FWHMs for
the $\gamma$ Dra data are:
5\farcs75$\pm$0\farcs08$\times$5\farcs56$\pm$0\farcs04 with Position
Angle (P.A.) of 32\arcdeg$\pm$29\arcdeg\ at 70 \mm,
6\farcs85$\pm$0\farcs04$\times$6\farcs70$\pm$0\farcs04 with P.A. of
33\arcdeg$\pm$26\arcdeg\ at 100 \mm, and
11\farcs97$\pm$0\farcs16$\times$10\farcs69$\pm$0\farcs15 with P.A. of
91\arcdeg$\pm$5\arcdeg\ at 160 \mm. The beam (defined as the FWHM) is
more elongated in the 160 \um channel and with a preferred
P.A. compared with the beams at 70 and 100 \mm. The variations in the
measured FWHMs are all less than 2\%, demonstrating the effectiveness
of the new gyro correction.

The final PACS images are shown in the upper panel of Figure
\ref{pacs_im}.  The measured FWHMs for the HD 95086 PACS data are:
7\farcs87$\times$7\farcs47 with P.A. of 112\arcdeg\ at 70 \mm,
8\farcs91$\times$8\farcs43 with P.A. of 116\arcdeg\ at 100 \mm, and
13\farcs4$\times$12\farcs2 with P.A. of 140\arcdeg\ at 160 \mm,
suggesting the source is resolved at $\sim$1.4 beams at both 70/100
\mm, but only at $\sim$1.1 beams at 160 \mm. This explains why the
P.A. measured at 160 \um is quite different from the the ones at
70/100 \mm. Based on the resolvability ($\lesssim$1.4 beam), we
simulated model images at all three bands (for details see below) and
at different inclinations, compared with the derived FWHMs and aspect
ratio, and concluded that the source is inclined at
25\arcdeg$\pm$5\arcdeg\ at P.A. of 115\arcdeg$\pm$10\arcdeg. These
values agree with the numbers derived by \citet{moor13}.

\begin{figure} 
  \figurenum{4}
  \label{SB}
  \epsscale{1.15}
  \plotone{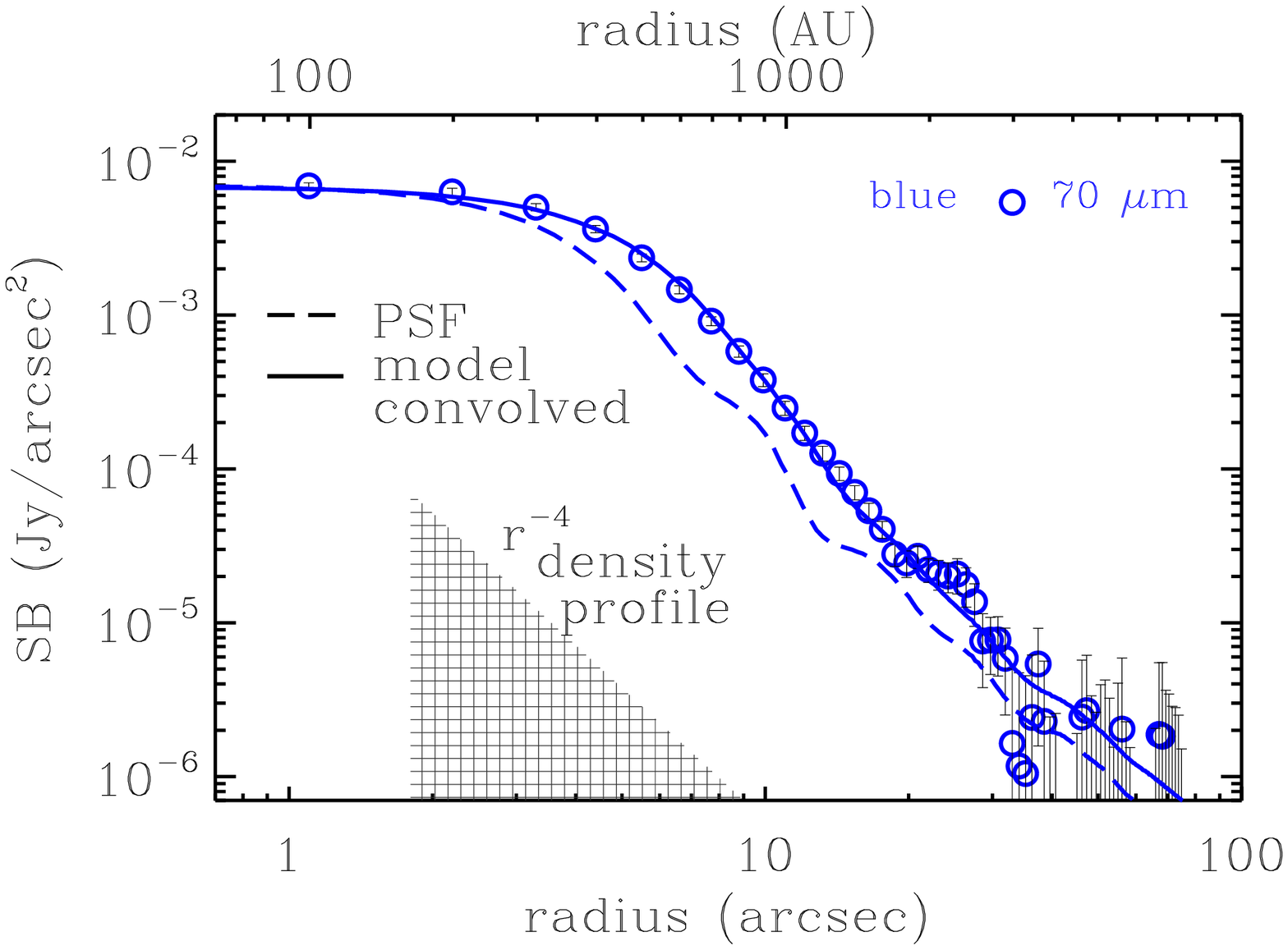}
  \plotone{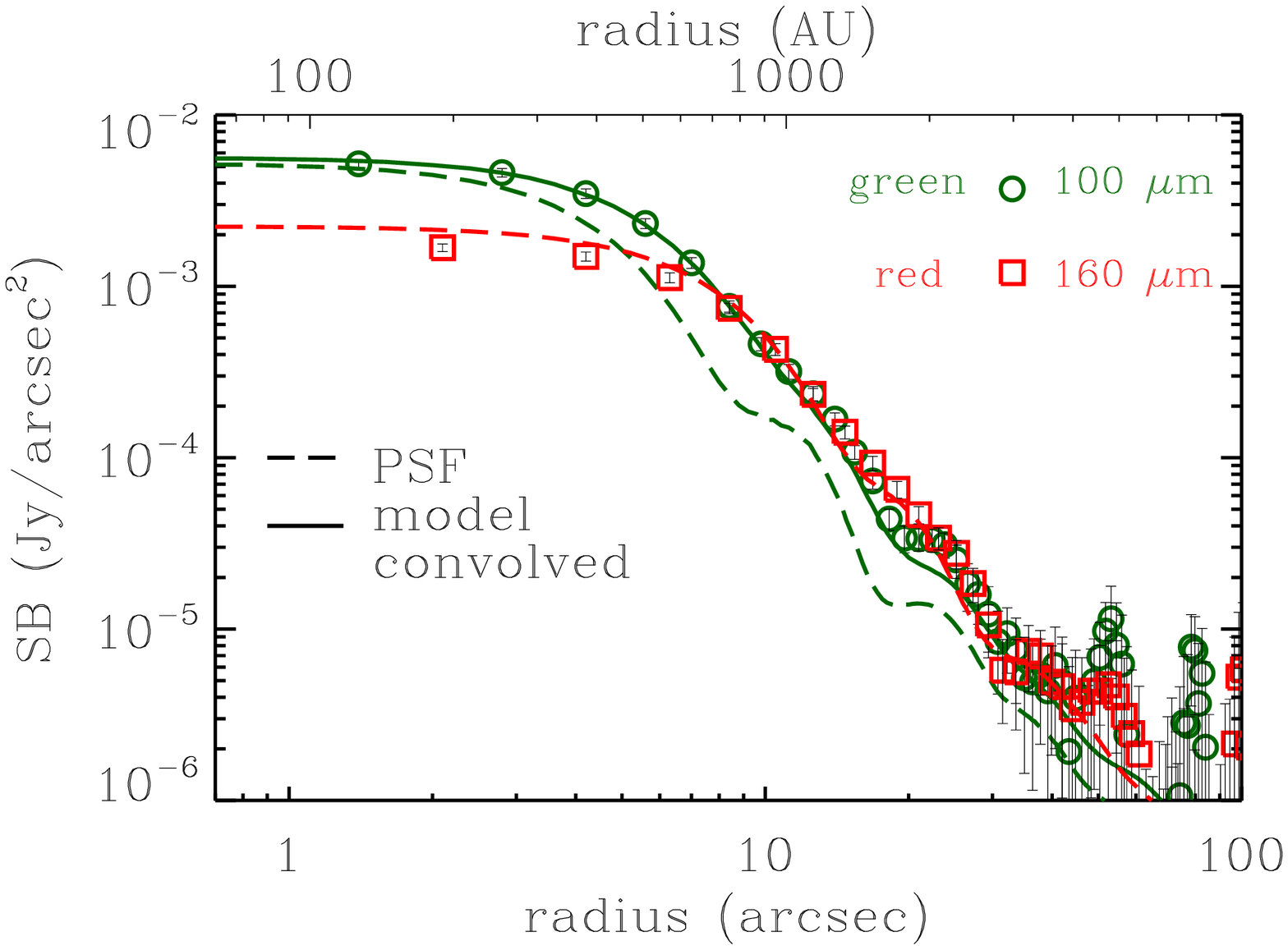}
  \caption{The azimuthally averaged radial surface brightness (SB)
profiles for HD 95086 (open circles: blue for data at 70 \mm, green
for data at 100 \mm, and red for data at 160 \mm). The stellar
contribution is negligible in the far-infrared. The radial profile of
$\gamma$ Dra (point source calibrator) is shown in dashed lines. Solid
lines shows the best-fit single power-law SB model after being convolved
with the PSF while the hash region on the upper panel illustrates the
model parameters ($\alpha=$4, $r_{in}=$1\farcs8, and
$r_{in}=$9\arcsec; details see the text). }
\end{figure}

The source is resolved at $\lesssim$1.4 beams, consistent with the
observation that the images at all three wavelengths are centrally
peaked. To estimate the maximum flux of a point source in the data, we
performed PSF subtraction using the $\gamma$ Dra combined PSF by
matching the peak value at the center.  The PSF-subtracted images are
shown in Figure \ref{pacs_im} bottom row.  The scaled point source
fluxes are 397 mJy, 422 mJy, and 357 mJy at 70, 100, and 160 \mm,
respectively, which are $\sim$262, $\sim$574, and $\sim$1266 times the
expected photospheric values at these wavelengths. The scaled point
source fluxes are also much brighter than the expected values of the
warm component ($<$10 mJy, see Figure \ref{exesed}). It is clear that
a significant amount of dust emission (presumably from a planetesimal
belt, P.B.) is not resolved (up to $\sim$58\%, $\sim$61\%, and
$\sim$73\% at 70, 100, and 160 \mm, respectively); i.e.,
$\gtrsim$30--40\% of the total excess fluxes is in the resolved
structure.

\begin{figure} 
  \figurenum{5}
  \label{radius_td} 
  \epsscale{1.15}
  \plotone{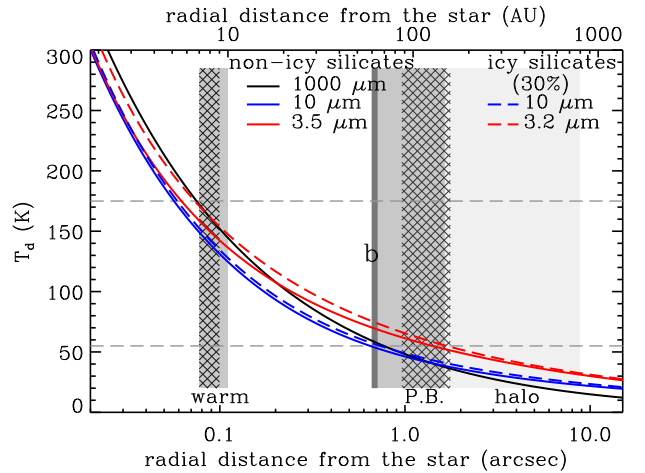}
  \caption{The dust temperature distribution for both astronomical
silicates and icy silicates at selected grain sizes. The two
horizontal dashed lines mark the two observed dust temperatures from
the disk SED. The location of the HD 95086b is marked at 62 AU. 
The ranges of the flat disk models are marked as dark gray areas while
the ones of Gaussian ring models are marked as hashed areas for both the warm belt
and cold planetesimal belt (P.B.). The light gray area shows the range of 
the disk halo. }
\end{figure}

\begin{figure*} 
  \figurenum{6}
  \label{modelsed} 
  \epsscale{1.15}
  \plotone{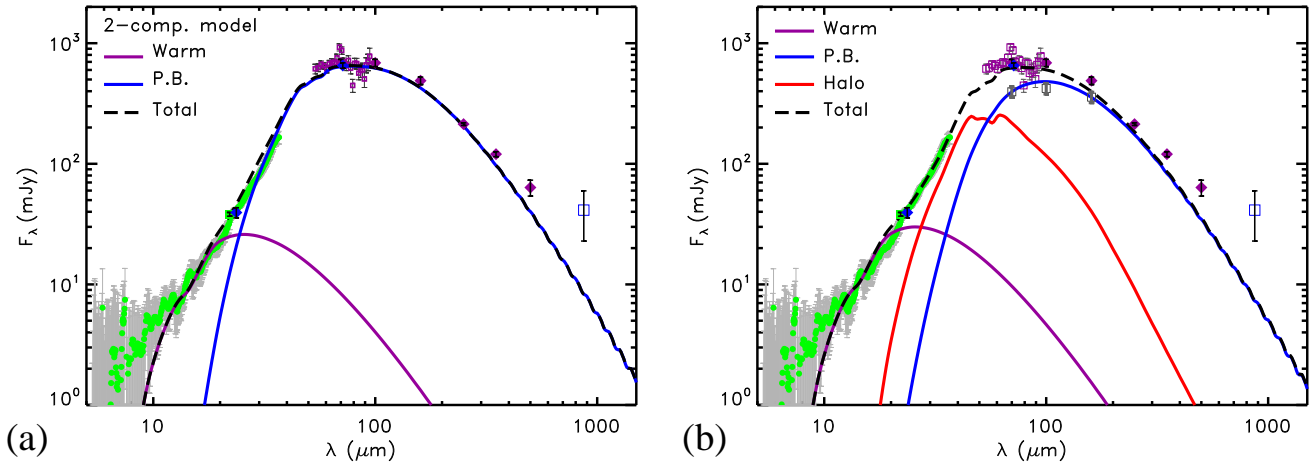}
  \caption{The spectral energy distribution (SED) of the debris around
HD 95086. In both panels, various points show the infrared excesses
after the removal of the stellar photosphere (symbols are used the
same way as in Figure 1), except that MIPS-SED data are shown as 
purple squares with grey error bars. Panel (a) shows the SEDs for
the two-component model. The infrared excesses can be
described with two separated flat disks: warm disk (purple
thick line: $R_{in}=7$ AU and $R_{out}=10$ AU) and planetesimal disk
(thick blue line: $R_{in}=63$ AU and $R_{out}=189$ AU). Panel (b)
shows one of the possible three-component models that provide a good fit
to the overall SED and resolved images.  Symbols used are the same as
panel (a) except for the new open squares representing the maximum
fluxes of the planetesimal belt at the PACS wavelengths (details see the
end of Section 3.2). }
\end{figure*}

We adopted a similar strategy as in \citet{su09} and
\citet{matthews14} to investigate simple structural parameters that
would be consistent with the resolved images. The model images were
constructed for a given model description, projected to the
inclination angle of 25\arcdeg, then convolved with the appropriate
PSF. We then computed the azimuthally averaged radial profiles for
both the observed and model data for comparison. Since PACS 70 \um and
100 \um channels have the better resolutions, we derived the best-fit
parameters only for the 70/100 \um data, and verified them with the
160 \um data. The model surface brightness (SB) was normalized to the
observed profile for data points within 7\arcsec\ (ensuring good fits
in the high signal-to-noise region), then we computed the $\chi^2$
values for data points within 20\arcsec\ at 70 \mm, and 15\arcsec\ at
100 \mm, respectively. Limited by the marginally resolved images, we
only explored a simple power-law distribution ($SB(r)\sim
r^{-\alpha}$) with inner ($r_{in}$) and outer ($r_{out}$) radial
cutoffs. With $\alpha=$3, we found $r_{in}\sim$1\farcs3 and
$r_{out}\sim$9\arcsec. With $\alpha=$4, we found $r_{in}\sim$1\farcs8
and $r_{out}\sim$9\arcsec. Alternatively, we could also find
reasonable fits for the case of $\alpha=$1, which is the power index
for the planetesimal belt derived for the resolved images of HR 8799
\citep{matthews14}. In summary, the power-law index ($\alpha$) is
unconstrained, and the resolved images are best described as a point
source plus some form of extended structure. As we discuss in Section
3.2, the resolved structure in the HD 95086 system most likely arises
from the disk halo component composed of only small grains that can be
warm enough to emit efficiently at far-infrared wavelengths to account
for the extended structure seen in the PACS images. We note that the
disk halo component is found to have $\alpha=$3--4 for HR 8799
\citep{matthews14}.  With this assumption, the derived $r_{in}$
($\sim$120--165 AU) represents the boundary where the halo emission
becomes dominant (i.e., the outer radius of the planetesimal belt),
while the $r_{out}$ ($\sim$800 AU) only represents the boundary where
the faint disk halo can be well detected given the sensitivity. The
best-fit SB distributions are shown in Figure \ref{SB} along with the
observed profiles. Unlike the HR 8799 SB profiles where two power-laws
were required to fit the data, one single power-law model provides
satisfactory fits to the HD 95086 data.

\subsection{SED modeling and Derived Parameters}

To guide the SED modeling, we first computed the thermal equilibrium
dust temperature distribution around HD 95086 (Figure \ref{radius_td})
with assumed grain properties in an optically thin environment. We
adopted two grain compositions: astronomical silicates and ice-coated
silicates. We used the optical constants from \citet{laor93} and Mie
theory code to compute the grain absorption and scattering efficiency
for compact, spherical astronomical silicates. For the absorption and
scattering efficiency of icy grains, we used both multi-layer Mie
calculations as well as standard effective medium theory rules
(Garnett and Bruggeman equation) (e.g.,
\citealt{voshchinnikov05}). For the inner warm belt, we adopt compact
astronomical silicates as the composition for the SED fitting since
the majority of the ice should be sublimated at the warm belt
temperature. For the outer cold disk, we use the icy grains composed
of compact silicate cores with icy mantles that occupy 30\% of the
grain volume. The thickness of the icy mantles is chosen so that the
resultant model emission produces a good overall fit in the 60--70 \um
region. Figure \ref{radius_td} shows that in terms of these two grain
compositions (non-icy vs.\ icy) the resultant temperatures for grains
larger than $\sim$10 \um are not very different, but they could have
$\gtrsim$10 K difference for \mm-size grains closer in. The observed
two characteristic temperatures ($\sim$175 K and $\sim$55 K) suggest
two distinct dust locations ($\sim$7 AU and $\sim$60 AU) for
blackbody-like grains.

Since fine debris is generated through collisional cascades by
breaking up large parent bodies, we expect a full spectrum of particle
sizes in a debris disk, generally in a power-law form, $n(a)\sim
a^{-q}$ with $n$ being the number density, $a$ for the grain radius
and $q=$ 3.5--3.7 \citep{wyatt11,gaspar12}. Thus, the majority of
particles in a collisional dominated disk are grains just slightly
larger than the minimum grain size in the distribution (i.e., the
average grain size is $<$$a$$>\sim \frac{q-1}{q-2}\ a_{min}$ where
$a_{min}$ is the minimum size of grains existing in a steady-state
collisional disk).  In a debris disk (no/very little gas), in addition
to the gravitational force of the star, dust grains are subject to
radiation pressure. Therefore, grains that are smaller than the
blowout size ($a_{bl}$) are removed from the system, i.e.,
$a_{min}\sim a_{bl}$. The blowout size is a function of stellar mass
($M_{\ast}$), luminosity ($L_{\ast}$) and grain density ($\rho$) given
by $a_{bl}=1.15 \frac{L_{\ast}}{L_{\odot}} \frac{M_{\odot}}{M_{\ast}}
\frac{g/cm^3}{\rho}$ in units of \mm.  Given the stellar parameters
for HD 95086 ($L_{\ast}$$\sim$7 $L_{\sun}$ and $M_{\ast}$$\sim$1.6
$M_{\sun}$), the radiation blowout radius is 1.5 and 1.8 \um
for densities of 3.3 $g\ cm^{-3}$ (for compact silicates) and 2.6 $g\
cm^{-3}$ (for icy silicates), respectively. Therefore, the average
grain sizes in such a collisionally dominated disk are $\sim$3 \mm. 
We also set the maximum grain radius
($a_{max}$) to be 1000 \mm; grains larger than this size have
negligible contribution to the model emission.

\begin{figure*} 
  \figurenum{7}
  \label{resim} 
  \epsscale{1.15}
  \plotone{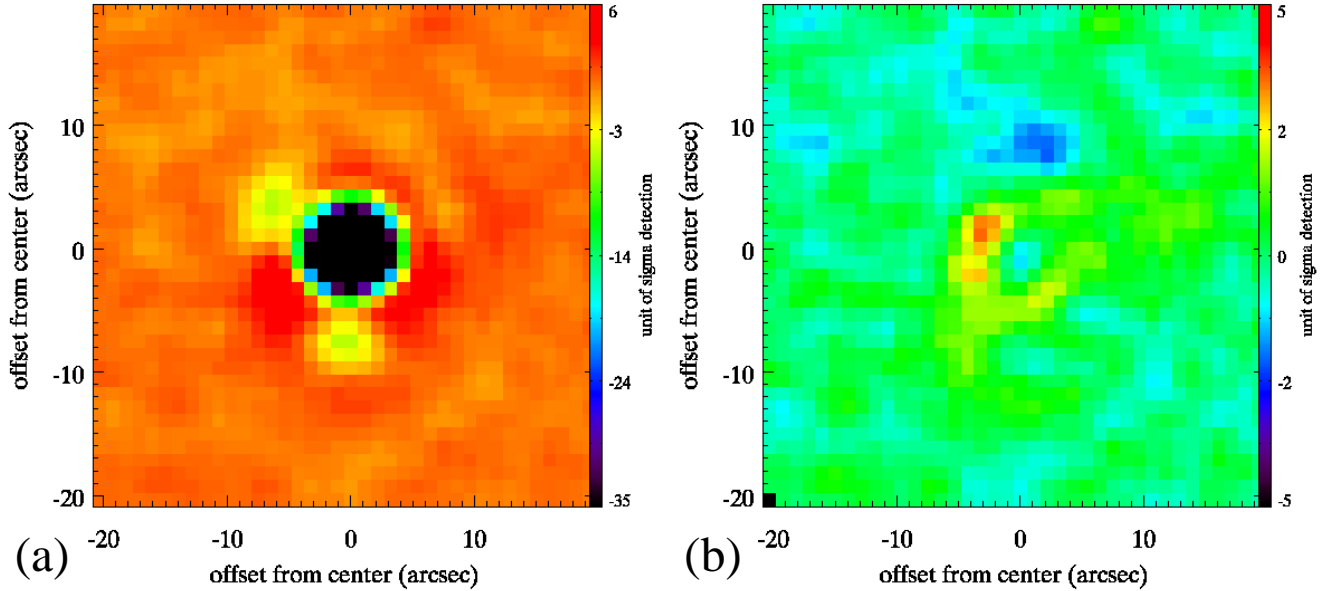}
  \caption{Difference images at 70 \um between observed and model
images: panel (a) using the two-component model while panel (b) using
the three-component model. The over-subtraction (more than
$-$100$\sigma$ at the center) in (a) is due to the fact that the
two-component model has no disk contribution outside the typical beam
(i.e., not resolved). In the three-component model, some portion of
the disk flux is distributed outside the beam (i.e., resolved disk
halo), and the residuals for the three-component model are within
$\pm$2$\sigma$. }
\end{figure*}

In our simple SED model, we adopt a uniform particle size distribution 
at all astro-centric distances and do not consider the effect of
grain size segregation due to stellar radiation pressure. The
grain-size segregation in debris disks is closely tied to the complex
coupling between dynamical and collisional effects (e.g.,
\citealt{thebault14}); therefore, the impact of using a uniform size
distribution is difficult to evaluate without multi-wavelength
resolved images. For simplicity, we adopt $q=3.5$ for the cold disk
component because the cold disk, as we show later, is quite
extended ($\sim$63--190 AU), and the particle distribution for a
unperturbed wide disk is most likely to have $q=3.5$ distribution
based on the computation of \citet{thebault14}. For the inner warm
component, on the other hand, we adopt a steeper power law, $q=3.65$,
based on the hypothesis that the warm component is a narrow belt
closer to the star where collisional cascades are expected to reach
collisional equilibrium \citep{gaspar12,thebault14}.  Although the
output far-infrared and submillimeter fluxes would be slightly
different depending on the $q$ values, the adopted $q$ value has
little impact on our model parameters for the warm component since we
have no far-infrared/submillimeter constraint.

Two kinds of simple density distributions were used to construct the
disk SED. One is a flat disk (constant surface density) with adjustable
inner ($R_{in}$) and outer ($R_{in}$) radii, and the other one is a
Gaussian profile ring with a peak radius ($R_{p}$) and a width
($R_{w}$ as the FWHM). With the grain parameters (composition and size
distribution) set, we searched for the best-fit two parameters in each
density distribution. Because the warm and cold components are not
spatially resolved and the (expected) peak emission of the warm
component (see Figure \ref{exesed}) is in the region where emission
from the cold component also contributes significantly, our SED
fitting strategy is to obtain reasonable fits to the cold component
first, then derive parameters for the warm component. For the
cold component, we only used the data points longward of 35 \um and
shortward of 500 \um to derive the best-fit (in $\chi^2$ sense) 
size parameters. In the flat disk model, the best-fit disk extension
is: $R_{in}=63\pm6$ AU and $R_{out}=189\pm13$ AU; while the Gaussian
ring model gives $R_p=123\pm3$ AU with a width of $R_w=72\pm11
AU$. For the cold disk, the derived dust fractional luminosity is
$\sim$1.5$\times10^{-3}$ with dust mass (up to 1000 \mm) of
$0.18\pm0.01 M_{\earth}$ for the flat disk, and $0.23\pm0.01
M_{\earth}$ for the Gaussian ring model.  Our dust masses are lower
than the value ($0.5\pm0.1 M_{\earth}$) derived by \citet{moor13} who
only used the 500 \um flux for mass estimation. As described in
Section 2.1, faint background galaxies within 40\arcsec\ of the source
are likely to contaminate the submillimeter fluxes. A better estimate
of the disk mass will have to come from high spatial resolution data
like those from ALMA.

We used non-icy grains for the warm component with $a_{min}=2.2$,
$a_{max}=1000$ \mm, and $q=3.65$. Because the warm component is less
constrained, the uncertainty from the SED fitting is large. Assuming a
flat disk density distribution, the warm component with radii ranging
from $\sim$7 to $\sim$10 AU gave a consistent fit to the shape of the
IRS spectrum. Alternatively, a Gaussian ring centered at 8 AU with a
width of 2 AU resulted in a similar fit. The derived dust fractional
luminosity is 1.5$\times10^{-4}$ with a total dust mass of
$\sim$4--5$\times10^{-5} M_{\earth}$. The model SED of the two (warm
and cold) flat disks is shown in Figure \ref{modelsed}a.

Although the two-component model gives an overall good fit to the
observed SED, the model images using these parameters do not fit the
observations. Figure \ref{resim}a shows the difference image
(observation -- model) for the two-component disk model at 70 \mm,
where very negative values at the source position are evident. The
extension of the cold disk derived from the observed SED (radius
$\lesssim$200 AU) implies that the cold disk should not have been
resolved with {\it Herschel}'s resolution (FWHM of 5\farcs6 at best).
We can obtain reasonable SED fits to the cold disk by extending it up
to $\sim$800 AU; however, the surface density distribution has to have
a steeper drop-off ($r^{-1.5}$ compared to a flat disk of
$r^{0}$). This very wide planetesimal disk, perhaps analogous to the
Solar System's scattered disk of KBOs, could generate dust grains
in-situ from collisions. Using a uniform particle size distribution
($a^{-3.5}$) at all astro-centric distances up to $\sim$800 AU, the
resultant model images from such a wide disk with a steep radial
gradient of surface density ($\sim$$r^{-1.5}$) are similar to the ones
using much smaller outer disk radii, i.e., not resolved. In order to
account for the detected far-infrared emission at large astro-centric
distances up to $\sim$800 AU, an extended component composed of only
small dust grains on unbound or barely bound orbits (i.e., a disk
halo) is our best alternative. These small grains are most likely
generated in the cold belt which may include a very wide scattered
disk structure. However, given that the source is only resolved at
$\lesssim$1.4 beams, no meaningful constraint can be placed based on
the images.

Without some information on the brightness of the planetesimal belt,
there is a wide range of parameters in this three-component SED model
that can produce satisfactory fits. For example, if we assume the
contribution of the planetesimal belt in the SED is roughly equal to
the point source fluxes that are scaled to match the peak fluxes of
the PACS images for subtraction, we are able to construct a
three-component model that fits the overall SED (Figure
\ref{modelsed}b) and the far-infrared images. In this three-component
model, the warm component is assumed to be the same as the
two-component model (it has negligible contribution at far-infrared),
and the emission at far-infrared is partitioned into the planetesimal
belt (blue line in Figure \ref{modelsed}b) and disk halo (red line in
Figure \ref{modelsed}b). The planetesimal-belt component has a dust
fractional luminosity of $\sim$8$\times10^{-4}$ and a total dust mass
of $\sim$0.2$M_{\earth}$ while the disk halo component has one third
($\sim$3$\times10^{-4}$) of the fractional luminosity in the
planetesimal-belt component, but only $\sim$10\% of the dust mass as a
result of the disk halo being composed of $\sim$\um size
grains. Figure \ref{resim}b shows a detailed comparison between the
observed and three-component model images at 70 \mm. Overall, the
residual suggests the three-component model agrees with the
observation within $\pm$2$\sigma$.

\section{Discussion}

\subsection{The Nature of the Warm Component}

Due to the large distance of HD 95086, the warm excess is inferred
from the SED only, unlike for other nearby debris systems (Vega and
Fomalhaut \citep{su13}, and $\epsilon$ Eri \citep{backman09}) where
the warm and cold components are spatially separated. Recently,
\citet{kennedy14} presented an informative way to examine the
properties of two-temperature disks, and tested the hypothesis that
emission from a single narrow belt can account for the two observed
temperatures by varying the particle size distribution. They conclude
that the one-narrow-belt scenario is unlikely to produce two observed
dust temperatures around early-type stars if the blowout size is
preserved.  Furthermore, the derived temperature and fractional
luminosity ratios between the HD 95086 warm and cold components
($R_T\sim 3.2$ and $R_{f_d}\sim$0.1--0.2) are located right at the
center of the $R_T$ vs. $R_{f_d}$ plot (Figure 5 in
\citealt{kennedy14}), indicating that varying the particle size
distribution in a narrow belt is unlikely to explain the observed SED.

Another possibility is that the warm excess emission could result from
the small amount of particles being dragged in due to P-R drag. The
effect of P-R drag in collisional dominated debris disks has been
studied analytically by \citet{wyatt05}. In the case of the HD 95086
disk (a planetesimal belt at $\sim$100 AU with $f_{d,cold}\sim10^{-3}$
producing excess emission at $\sim$10 AU), the maximum amount of
material available due to P-R drag in terms of optical depth is
$<$0.05$f_{d,cold}$. Furthermore, the analytical maximum value was
found to be a factor of $\sim$7 higher than using a numerical model
with a better treatment of collisional effects
\citep{vanLieshout14}. Thus, the observed value
$f_{d,warm}\sim$1.5$\times10^{-4}$ for the warm component, is a factor
of 20 higher than the maximum value that the cold belt can supply due
to P-R drag, making this hypothesis unlikely. We, therefore, consider
the warm component as an independent belt in the following discussion.

\subsection{Detection of the 69 \um Forsterite Feature} 

The 69 \um emission band, a solid state feature originating from
magnesium-rich olivine crystals, was first observed astronomically by
{\it ISO} around a Herbig Ae/Be star, HD 100546 \citep{malfait98}. The
peak position and strength (FWHM) of the 69 \um feature strongly
depends on the iron content of the crystals and their temperatures
\citep{bowey02,koike03,suto06}. See \citet{sturm13} and
\citet{maaskant14} for recent reviews on the iron-content and
temperature dependence, and other secondary effects like grain sizes
and shapes. The feature detected in the HD 95086 MIPS-SED data is not
spectrally resolved, we cannot put constraints on the fraction of iron
composition as has been done for $\beta$ Pic \citep{devries12};
however the iron fraction has to be less than a few \% to have the
feature peaked at 69 \um at $\sim$50 K.

To further validate the detection of the forsterite
69 \um feature and derive a rough estimate of the crystalline abundance,
we took the mass absorption coefficients of synthetic forsterite
(Fo100 in \citealt{koike03}), and constructed simple model fits to the
MIPS-SED data. The model has two components: (1) emission from
the crystalline forsterite and (2) a blackbody emission representing
the underlying dust continuum, with both at the temperature of 55 K
derived for the cold component. The model spectrum was smoothed to
match the MIPS-SED resolution, and the ratio between the two
components was adjusted to fit the overall shape of the MIPS-SED
data. Since we do not have information about the location
where the 69 \um feature originates, the temperature of this component
is not necessarily the same as that of the cold disk (underlying
continuum). However, we can easily rule out the possibility that 
these crystalline
grains originate from the warm belt with a dust temperature of
$\sim$175 K since the blue part (50--60 \mm) of the model spectrum
would have a much bluer slope. Given that the continuum in both blue
and red sides of the feature is relatively flat, we assume the
crystalline grains are located in the cold disk with the same dust
temperature as the underlying continuum.

\begin{figure} 
  \figurenum{8}
  \label{fo100} 
  \epsscale{1.15}
  \plotone{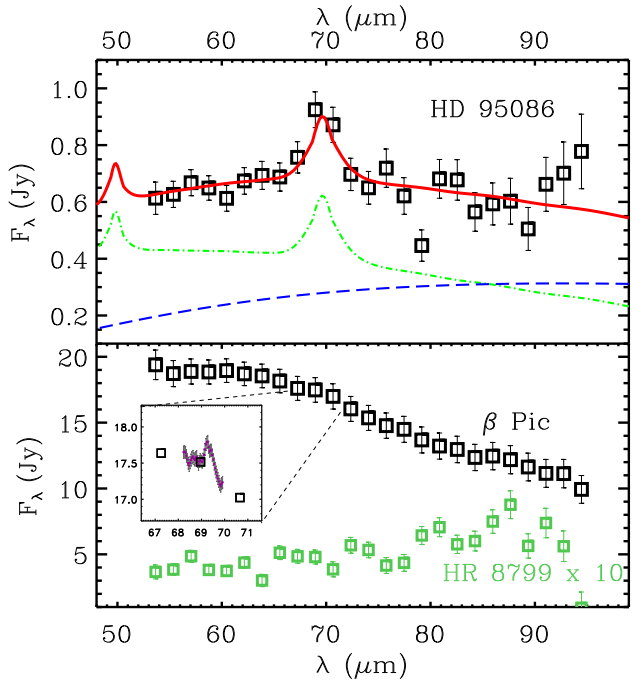}
  \caption{Photosphere-subtracted MIPS-SED data (open squares with
error bars) for HD 95086 (upper panel), $\beta$ Pic and HR 8799
(bottom panel). Our best-fit model, shown as the solid red line in the
upper panel, consists of the underlying dust continuum (55 K blackbody
emission, the blue dashed line) and the emission of crystalline
forsterite (the green dotted-dashed line) after smoothing to match the
MIPS-SED resolution. The MIPS-SED data of HR 8799 in the bottom panel
was scaled by 10 times for easy comparison. The inset of the bottom
panel shows the scaled (by 1.07) PACS spectroscopic data (purple line) from
\citet{devries12}.}

\end{figure}

The upper panel of Figure \ref{fo100} shows the best-fit model where
the total mass of the forsterite component is $\sim$6.2$\times10^{25}$ 
g (0.01 $M_{\earth}$). Using the total dust mass we derived from the
cold disk (0.2 $M_{\earth}$, Section 3.2), the abundance of the
crystalline grains is $\sim$5\%.  This abundance is likely to be a
lower limit because the laboratory measured absorption coefficients
are based on finely ground powder (sizes of 0.1--0.5 \mm), and the 69
\um feature is not sensitive to grain size (the band profiles are
similar for grains less than $\sim$10 \mm,
\citealt{sturm13})\footnote{The changes in the peak position and FWHM
of the 69 \um band profiles at different grain sizes are all much
smaller than the spectral resolution of the MIPS-SED mode
data.}. Since no prominent crystalline features are seen in the
observed 20--40 \um range (while our best-fit model predicts strong
features), the crystalline grains must have sizes $\gtrsim$ a few
\mm. Therefore, the derived abundance should be taken as a lower
limit.

The presence of crystalline silicates at a low temperature and far
from the star is surprising, given that relatively high temperatures
are required to convert circumstellar amorphous silicates to the
crystalline form \citep{fabian00}. The detection of the forsterite
emission feature implies that there is a substantial transport of
material from the inner zone of the system, or that material has been
released from the disruption of a large body where high temperatures
were reached during formation.

It is interesting to note that the 69 \um feature was not detected in
the MIPS-SED data of $\beta$ Pic (the bottom panel of Figure
\ref{fo100}), while the crystalline abundance is estimated to be
3.6$\pm$1.0\% using the PACS spectroscopic data
\citep{devries12}. With the published PACS spectrum smoothed to match
the MIPS-SED resolution, we found that the low equivalent width of the
detected feature is completely washed out in the MIPS-SED mode (the
inset of the bottom panel of Figure \ref{fo100}).  This is also
consistent with the derived mass values for the crystalline grains
(2.8$\times10^{23}$ g for the $\beta$ Pic disk, while the mass we
estimate for HD 95086 is $\gtrsim$200 times more).  The estimated
crystalline abundances implied by the observed feature strengths in
these two systems appear discrepant. The discrepancy lies in the total
dust mass used to derive the abundance, which is model dependent. The
SED models used for the $\beta$ Pic system include grains up to 100
\um for all three-temperature components \citep{devries12}, while our
models for HD 95086 include grains up to 1000 \mm, i.e., a factor of
$\sim$3 in the derived dust mass in a typical $a^{-3.5}$ size power
law ($M_d\sim \sqrt{a_{max}}$). Indeed, the total non-crystalline,
derived dust mass by \citet{devries12} is a factor of $\sim$3 lower
than the value (4.7$\times10^{26}$ g) derived from recent ALMA
observation for $\beta$ Pic \citep{dent14}. Furthermore, the reported
3.6\% crystalline abundance is only based on one of the temperature
components ($\sim$90 K with a total mass of 8$\times10^{24}$ g). Using
the ALMA derived dust mass, the crystalline abundance in the $\beta$
Pic disk should be $\lesssim$1 \%. As we compare the disk properties
between HD 95086 and HR 8799 in the following subsection, it is also
noteworthy that no feature was detected in the MIPS-SED data around HR
8799 (Figure \ref{fo100}) \citep{su09}, suggesting the crystalline
abundance is $<$1 \%.

\subsection{Constraints on the Eccentricity of the Disk}

The eccentricity, $e$, of debris disks usually can be measured in two
ways: (1) the offset between the ring center and star position, and
(2) the degree of asymmetry seen in thermal images (i.e., pericenter
glow). Both methods have been applied to the Fomalhaut debris disk
\citep{kalas05,stapelfeldt04} and result in a consistent value
($e\sim0.1$). Since the star contributes negligibly in the PACS
wavelengths, the centroid we measured from the PACS images is the
center of the disk. Based on the calibration observations of the PSF
star, $\gamma$ Dra, a typical centroiding uncertainty is
0\farcs01. The position of the star is well measured by 2MASS with a
known proper motion from Hipparcos; therefore, the uncertainty in its
position relative to the {\it Herschel} detection is dominated by the
absolute accuracy of the World Coordinate Systems (WCS) in {\it
Herschel}/PACS maps. We used a bright galaxy (LEDA 3079352, 2MASS
J10571665-6838171) present in the field of our PACS images to estimate
the absolute pointing uncertainty. There appears to be an offset of
$\sim$2\farcs2 between the 2MASS and {\it Herschel} coordinate
systems. The typical pointing uncertainty in {\it Herschel}
observations ($\sim$2\arcsec) makes it difficult to pin point the star
position with high accuracy (i.e., the first method is inapplicable).

We then turned to the second method to assess the degree of asymmetry
we could have detected with {\it Herschel} observations. Since the
resolved structure is mostly dominated by the disk halo (the
planetsimal belt is not resolved at {\it Herschel}'s resolution), the
presumption is that the disk halo shape follows closely with that of
the planetesimal belt. This assumption is reasonable if the disk halo
results from small grains generated in the parent-body belt that
are ejected and/or placed in highly eccentric orbits due to radiation
pressure. There appears to be some asymmetry seen in the
PSF-subtracted images (bottom row of Figure \ref{pacs_im}); however,
it is not significant (less than 3$\sigma$). The resolved component is
consistent with the circular case with $e=0$ (our baseline model shown in
Figure \ref{modelsed}b and Figure \ref{resim}b). We then constructed
model images by placing the disk halo components at various
eccentricities and compared them with our baseline model. We found that
the difference is noticeable (more than 5$\sigma$) when $e>0.25$. From
this exercise, we conclude that the resolved structure (disk halo)
around HD 95086 has $e<0.3$.

\subsection{Is HD 95086 A Young Analog of HR 8799? }

Given that both HR 8799 and HD 95086 have bright debris disks and
wide-orbit planets revealed via direct imaging, it is instructive to
compare their properties in regard to different aspects of the
planetary configuration. Table \ref{tbl:comp} lists the properties of
the HR 8799 and HD 95086 systems. The debris structures revealed from
excess SEDs and resolved images show that both systems have a warm
($\sim$170 K) excess near the water-ice line and a cold excess at
$\sim$50 K, surrounded by an extended disk halo. The HD 95086 system
is $\sim$10 times dustier than the HR 8799 in terms of fractional dust
luminosity, but the ratio of fractional luminosities of the warm and
cold components is similar ($\sim$0.1). The extended disk halo in HR
8799 can be traced up to $\sim$2000 AU with a sensitivity of 10$^{-2}$
mJy arcsec$^2$ \citep{matthews14}; interestingly, a similar size is
seen in the HD 95086 system with the same detection limit, implying
its disk halo is $>$5 times brighter given that it is $\sim$2.3 times
more distant. Based on the resolved images, \citet{matthews14}
estimated the HR 8799 disk halo accounts for 40\%, 46\% and 52\% of
the total flux at 70, 100 and 160 \mm, respectively. In the case of HD
95086, the marginally resolved images can only give a lower limit for
the contribution from the disk halo, which is $\gtrsim$42\%,
$\gtrsim$39\% and $\gtrsim$27\% at 70, 100 and 160 \mm,
respectively. In other words, the debris structures in these two
systems are very similar except that the debris in the HD 95086 system
is $\sim$10 times more in dust luminosities and 2--5 times more in
observed dust masses (derived from SED fitting) compared to those
values in the HR 8799 system.

\begin{deluxetable}{ccccc}
\tablewidth{0pc}
\tablecaption{Properties Comparison between HR 8799 and HD 95086 \label{tbl:comp}}
\tablehead{ 
\colhead{}            & \multicolumn{2}{c}{HR 8799 } &  \multicolumn{2}{c}{ HD 95086} 
}
\startdata  
Stellar Parameter & & & & \\ 
$M_{\ast}$ [$M_{\sun}$] &  \multicolumn{2}{c}{1.5} &   \multicolumn{2}{c}{1.6} \\
$L_{\ast}$ [$L_{\sun}$] &  \multicolumn{2}{c}{5.7} &   \multicolumn{2}{c}{7.0} \\
$T_{eff}$ [K]           &  \multicolumn{2}{c}{7500} &  \multicolumn{2}{c}{7500} \\
$R_{\ast}$ [$R_{\sun}$]  & \multicolumn{2}{c}{1.4}  & \multicolumn{2}{c}{1.6} \\
age [Myr]                    &  \multicolumn{2}{c}{$\sim$30--90$^{[1]}$}   &  \multicolumn{2}{c}{$\sim$20$^{[2]}$}  \\
distance [pc]               &  \multicolumn{2}{c}{39.4}        &  \multicolumn{2}{c}{90.4}    \\ 
\hline 
\\
SED Parameter & & & & \\ 
                       &  {\it warm}    & {\it cold}   & {\it warm}   & {\it cold}  \\ 
$T_{d}$ [K]               &  $\sim$150    & $\sim$45    & $\sim$175   & $\sim$55   \\
$f_{d}$    &  1.8E-5   & 2.2E-4   &  1.5E-4   & 1.5E-3  \\ 
$M_{d}$ [$M_{\earth}$]    &  1E-6   & 0.1   &  5E-5   & 0.2--0.5  \\ 
\hline 
\\
Disk Orientation & & & & \\ 
$i$ [degree]   & \multicolumn{2}{c}{25$\pm$5}  & \multicolumn{2}{c}{25$\pm$5}  \\
P.A. [degree]  & \multicolumn{2}{c}{62$\pm$10}  & \multicolumn{2}{c}{115$\pm$10}  \\
\hline
\\
Planet &   \multicolumn{2}{c}{e, d, c, b}  & \multicolumn{2}{c}{b}  \\
$M_p$ [$M_J$]   & \multicolumn{2}{c}{9$\pm$2, 9$\pm$3, 9$\pm$3, 7$\pm$2$^{[3]}$}  & \multicolumn{2}{c}{5$\pm$2$^{[4]}$}  \\
$a_p$ [AU]   & \multicolumn{2}{c}{15.4, 25.4, 39.4, 69.1}  & \multicolumn{2}{c}{61.5$^{+5.7}_{-4.9}$$^{[5]}$}  \\
\enddata
\tablerefs{[1] \citet{baines12}, see Sec 4.4 for details; [2] \citet{meshkat13}; [3] \citet{gozdziewski14}; [4] \citet{rameau13b}; [5] this work, assuming $e_b$=0.}

\end{deluxetable}

That both systems are of 10--100 Myr age is exciting because this is
the epoch when interesting processes like dynamical settling,
formation of rocky planets, and possibly even (late) formation of ice
giants are expected to occur in planetary systems. The age estimate of
HD 95086 ($\sim$20 Myr) is quite robust given its association with the
LCC moving group; but the age of HR 8799 has been widely debated. Most
of the dating methods applicable for early-type stars favor an age
younger than $\sim$150 Myr \citep{marois08}, while asteroseismic
analysis based on the ground-based data favors an older age of $\sim$1
Gyr \citep{moya10}. Recent optical monitoring using the
Microvariability and Oscillations in STars (MOST) space data found
significant frequency and amplitude changes in HR 8799's pulsation,
making precise asteroseismic analysis difficult \citep{sodor14}. The
age of HR 8799 was revisited by \citet{baines12} with the directly
measured stellar diameter from interferometric CHARA data and
evolutionary models, who derived two age estimates: 33$^{+7}_{-13.2}$
Myr if the star is still contracting toward the zero-age main sequence
(ZAMS), and 90$^{+381}_{-50}$ Myr if the star is evolving from
ZAMS. Therefore, it is possible that HR 8799 could be as young as HD
95086 (20 Myr), but it is more likely older, by up to $\gtrsim$ 10
times.

If both HR 8799 and HD 95086 are at the same age, the difference in
the observed dust level must reflect different initial
conditions. Based on the evolution models of debris disks developed by
\citet{kenyon08}, the relative disk luminosity ($f_d$) around a 2
$M_{\sun}$ star at age of 20 Myr can be up to 10 times higher if one
disk started with 10 times more mass than the other (Figure 17 in that
paper). This suggests that HR 8799 was born with a less massive
protoplanetary disk than that of HD 95086 if both have the same
age. However, all four directly imaged planets in the HR 8799 system
are $\sim$2 times more massive than the HD 95086 planet, implying the
opposite trend in initial conditions. Alternatively, if both systems
formed with similar initial conditions (mass and angular momentum),
the difference in the dust level must be related to debris
evolution. The models by \citet{kenyon08} predict the decline in disk
luminosity is roughly a power law, $f_d \propto t^{-n}$ with $n
\approx$0.6--1.0. With possible ages of 40--200 Myr for HR 8799, the
differences in $f_d$ between HD 95086 and HR 8799 suggest
$n\sim$1.0--3.3, a steeper decline than the model prediction.

Perhaps the biggest difference between these two systems is that HR
8799 hosts four massive (7--9 $M_J$) planets between the warm and cold
belts, while there is only one $\sim$5 $M_J$ planet currently known
that is just interior to the HD 95086 cold belt. Because HD 95086 is more
than two times more distant than HR 8799, the current direct imaging
observations are only sensitive to $\gtrsim$5 $M_J$ as close as
$\sim$30 AU to the star \citep{rameau13a}. Similar to the
well-resolved two-belt systems (HR 8799, Vega, Fomalhaut and
$\epsilon$ Eri), the large gap between the warm and cold excesses
might host multiple planets.

\subsection{Dynamical Constraints on Possible Planet Configurations}

We can constrain possible planet configurations of the HD 95086 system
based on two dynamical considerations: (1) multiple planets must be
dynamically stable for a time span of the age of the HD 95086 system,
$\sim$20 Myr, and (2) the planets' gravitational perturbations must
clear and maintain the gap between the warm and cold belts.  For the
former, we make use of the numerical results of \citet{faber07} on the
stability of multiple planet systems in coplanar configurations of
roughly geometrically spaced orbits.  For the latter we make use of
numerical estimates of a planet's chaotic zone in which debris can be
cleared and a gap can be maintained by each planet.  The chaotic zone
is owed to the overlap of first order mean motion resonances, as shown
by \citet{wisdom80}, who estimated the chaotic zone width as $\Delta a
\simeq 1.3 \mu^{2\over7}a_p$, where $\mu=M_p/M_*$, and $M_p, a_p$ are
the planet's mass and orbital semi-major axis.  In this zone,
initially circular test particle orbits exhibit strongly chaotic
motion.  Subsequent numerical studies have supported this estimate for
planet-to-star mass ratios, $\mu$, up to about $10^{-3}$
\citep{duncan89}. \citet{morrison14} carried out numerical
integrations for $\mu$ up to $10^{-1.5}$, to determine the range in
the planet's chaotic zone where particles are not merely chaotic, but
actually cleared (by either collision or escape); they report cleared
zone widths, both interior and exterior to a planet's circular orbit,
as follows: $\Delta a_{cl, int}=1.2\mu^{0.28}a_p$ and $\Delta a_{cl,
ext}=1.7\mu^{0.31}a_p$, In this section, we use these ``cleared zone''
widths to estimate the effects of planets on the clearing of debris.
However, we continue to refer to these as ``chaotic zone'' widths, to
acknowledge the underlying dynamical mechanism of clearing. Below, we
consider possible planetary systems in coplanar, non-resonant
configurations.

\subsubsection{Single Planet System}

Could HD 95086b alone be responsible for the gap, $\sim$8 AU to
$\sim$80 AU, between the warm and the cold belts?  HD 95086b's
estimated mass, $5\pm2M_J$, and its observed location imply that it
orbits within the gap and that it is at least shepherding the inner
edge of the cold belt.  The projected astro-centric distance of HD
95086b of $55.7\pm2.5$ AU and the disk inclination angle of
$25^\circ\pm5^\circ$ yield a de-projected distance,
$r_b=61.5_{-4.9}^{+5.7}$~AU. The separation of the planet from the
inner edge of the cold belt is $\sim$$(80-r_b)\approx(13$--23) AU.
For the nominal planet mass, $M_b=5M_J$, and nominal circular orbit of
radius $a_b\approx61.5$~AU, this separation distance is similar to the
expected range, $\Delta a_{cl,ext} = 1.7(M_b/M_*)^{0.31} a_b \approx
17 $~AU, of clearing of test particles in the planet's chaotic zone
exterior to its orbit.

To be responsible for the whole gap, the planet would have to be on a
highly eccentric orbit, with apocenter near its present location, and
pericenter near the warm belt (actually near one chaotic zone width
removed from the warm belt), i.e., with an orbital eccentricity
$e_b\approx0.7$.  Such a high eccentricity of the planet's orbit would
then force a high eccentricity on the debris belts too.  The
eccentricity of the debris belts are undetermined from the current
observations; therefore, it cannot be ruled out that HD 95086b on an
eccentric orbit may be the cause of the large gap between the warm and
cold belts.  However, it begs the question of how HD 95086b would have
gained such a highly eccentric orbit.  It is more plausible that
multiple planets exist in this large gap.

\begin{figure}[t] 
  \figurenum{9}
  \label{fig:plloc} 
  \epsscale{1.15} 
  \plotone{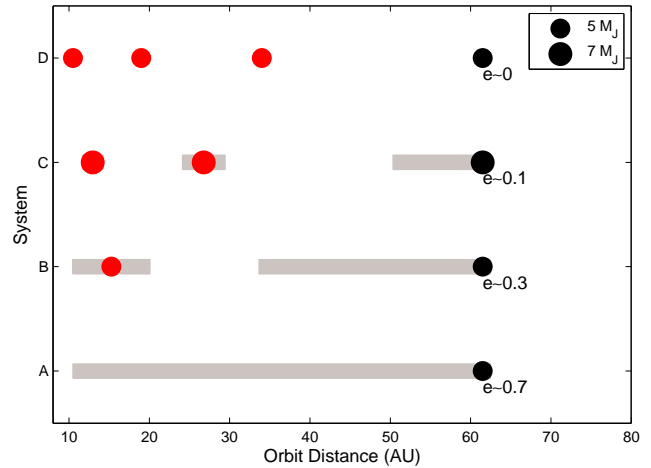}
  \caption{Possible planetary architectures of HD 95086 that are
dynamically stable or marginally stable for the age of the system and
that can also account for the gap between the warm and cold debris
belt.  The black points show the nominal location and mass of HD
95086b; red points indicate the semi-major axis of hypothetical
(unseen) planets; the grey bars indicate the possible range of
astro-centric distances of each planet, in the systems with
eccentric orbits. Size of points indicate planet mass.}
\end{figure}

\subsubsection{Two Planet System}

Let us consider whether the gap could be maintained by two planets on
low eccentricity ($e\sim$0.1) orbits. The known planet, HD 95086b, of
mass $\sim5\pm2~M_J$ on a nearly circular orbit of radius
$\sim61.5\pm5$ AU would both shepherd the outer belt as well as clear
debris inward of its orbit by about one chaotic zone width, i.e.\ down
to about 42 AU at most.  The second unseen planet must be located near
the inner belt. Assuming that this planet is of mass $\sim$13 $M_J$
(the conventional upper limit of planetary masses), and its pericenter
is located one chaotic zone width ($\Delta a_{cl,int}$) away from the
outer edge of the inner belt, we estimate that its semi major axis is
$\sim$13 AU. Such a putative planet could shepherd the inner
belt and clear debris out to a maximum distance of about 18
AU. This is insufficient to account for the entire 8--80 AU observed
gap.  We can also consider a higher mass, unseen sub-stellar object
near the inner belt.  Currently, the best detection limit ($5\sigma$)
for low-mass companions close to the star is $\sim21 M_J$ at a
astro-centric distance of 0\farcs18 ($\sim$16 AU) \citep{meshkat14}.
Demanding that a $21 M_J$ object be able to shepherd the inner belt
edge at $\sim$8 AU implies that its pericenter is located at about one
chaotic zone width away, i.e., at about 14 AU. This means that
it can clear debris out to about $(14 + \Delta a_{cl,ext})\approx
20$ AU.  This is also insufficient to account for the entire 8--80 AU
gap.  We conclude that two planets on low eccentricity orbits are
unable to account for the gap.

If we relax the low eccentricity condition, then it is possible to
account for the gap provided one or both planets' orbits are of
significant eccentricity.  First, let's consider that HD 95086b is of
low eccentricity but the inner unseen planet is of high eccentricity.
Then, as we noted above, HD 95086b can clear debris inward of its
orbit to about $\sim$42 AU astro-centric distance, so we require the
inner unseen planet to account for debris clearing in the range 8--42
AU.  Adopting the maximum mass for the inner planet, $M_1=13 M_J$, we
can estimate its pericenter and apocenter distances: $q_1\simeq
8/(1-1.2(M_1/M_*)^{0.28})\simeq$12 AU, and $Q_1 \simeq
42/(1+1.7(M_1/M_*)^{0.31})\simeq$30 AU.  Therefore, the inner
planet (of maximum planet mass $13 M_J$) must have orbital
eccentricity $\sim$0.5 to account for the observed gap.  Solutions
with somewhat lower planet mass and eccentricity are possible.  For
example, consider two planets of similar mass, $5M_J$, and similar
orbital eccentricity, $\sim$0.3, with the inner planet's orbit of semi
major axis $\sim$16 AU and HD 95086b's orbit of semi major axis
$\sim$40 AU.  Their pericenter and apocenter distances, augmented with
one chaotic zone width each, could shepherd the warm and cold belt
edges as well as clear the 8--80 AU gap; moreover, the planets' orbits
are also marginally stable for the 20 Myr age of the system. This
two-planet configuration is illustrated in Figure 9 as ``system B''.

\subsubsection{Three Planet System}

Given the young age of this system, it is plausible that the gap
contains multiple (more than two) planets in orbits that are only
marginally stable on $\sim$20 Myr timescales.  The parameter space of
such multiple-planet configurations is very large, of course.  To
obtain some rough estimates, we consider systems of three and four
planets, with equal-mass planets on coplanar low eccentricity
($e<$0.1) orbits, and demand stability for only $\sim$20 Myr. Then we
identify a few possible configurations as follows.

In a putative three-planet system, let's designate the three planets
with index 1, 2, 3 in order of increasing astro-centric distance.
Planet 1 is identified with the one that must shepherd the outer edge
of the warm belt (located at $\sim$8 AU), while planet 3 is identified
with the detected planet, HD 95086b (currently located at
astro-centric distance $r_b\approx61.5\pm5$~AU).  Then, the pericenter
distance of planet 1, $q_1=a_1(1-e_1)$, must be approximately one
chaotic zone width ($\Delta a_{cl,int}$) away from the edge of the
warm belt, i.e. $q_1\approx 8/(1-1.2\mu_1^{0.28})$~AU where
$\mu_1=M_1/M_*$ is the planet-to-star mass ratio.  Adopting the
1-$\sigma$ upper limit of the mass of planet b ($7~M_J$) for the
equal-mass planet 1, and $e_1=0.1$, we find that $a_1\simeq 12$~AU.
For planet b, assuming it is currently near apocenter and that it has
orbital eccentricity $e_b=0.1$, we find $a_3\simeq 56\pm5$~AU.  To
estimate the orbit of planet 2, let's assume that its fractional
separation from planet 1 is the same as from planet 3, i.e.,
$a_2=a_1(1+\delta)=a_3/(1+\delta)$; this yields $\delta\simeq 1.16$
and thereby $a_2\simeq26$~AU.  We then ask if this three-planet
configuration is dynamically stable for a time span of $\sim$20 Myr,
which is about $10^6$ orbital periods of the innermost planet.
Referring to the pertinent numerical study of \citet{faber07} who
investigated dynamical stability of multiple planets in coplanar low
eccentricity orbits of equal fractional separations, we find that
$\delta=1.16$ exceeds by a comfortable margin the minimum value,
$\delta_{{min}}\simeq 3.5\mu^{1/4}\simeq0.9$ for
$\mu=7M_J/1.6M_\odot=0.00418$, required for dynamical stability for a
time span of $10^6$ orbital periods of the innermost planet. Hence,
this three 7$M_J$ planet system is dynamically stable for $\sim$20
Myr.

We then ask if this planet configuration may account for the entire
8--80 AU gap.  By construction, planet 1 and planet 3 shepherd the
edges of the warm and cold belts, respectively.  We calculate that the
exterior chaotic zone of planet 1 (measured from its apocenter) just
marginally overlaps the interior chaotic zone of planet 2 (measured
from its pericenter); similarly, the exterior chaotic zone of planet 2
(measured from its apocenter) just overlaps the interior chaotic zone
of planet 3 (measured from its pericenter). Thus, within the
observational uncertainties, we consider that this configuration of
three equal-mass planets, each of mass $7M_J$, in low eccentricity
orbits of semi major axes $\sim$12 AU, 26 AU and 56 AU, is stable for
the age of the system and can also (marginally) account for the entire
gap between the warm and cold belts. This three-planet configuration
is illustrated in Figure 9 as ``system C''.  We note that three
planets of much smaller mass (or of vanishingly small orbital
eccentricities) would be less likely to meet the observational
constraint of maintaining the large gap.  Thus, we are motivated to
consider a four planet system of lower mass planets.

\subsubsection{Four Planet System}

In a putative four, equal-mass, we designate the four planets with
index 1, 2, 3, 4 in order of increasing astro-centric distance.  As
above, planet 1 is identified with the one that must shepherd the
outer edge of the warm belt (located at $\sim$8 AU), while planet 4 is
identified with the detected planet, HD 95086b (currently located at
astro-centric distance $r_b\approx61.5\pm5$~AU).  Assuming circular
orbits, planet 1 must be approximately one chaotic zone width ($\Delta
a_{cl,int}$) away from the edge of the warm belt, i.e. $a_1\approx
8/(1-1.2\mu_1^{0.28})$~AU; adopting the nominal mass of planet b
($5~M_J$) for the equal-mass planet 1, we find that $a_1\simeq$11 AU.
For planet b, we have $a_b\simeq(61.5\pm5)$~AU.  We assume that the
adjacent planets' fractional orbital separations are equal, i.e.,
$a_{j+1}=a_j(1+\delta)$.  We calculate $\delta = (a_b/a_1)^{1\over3}-1
\simeq 0.8$ and thereby $a_2\simeq $19 AU and $a_3\simeq$34 AU.
Again, referring to the results of \citet{faber07}, we calculate that
$\delta_{min}=3.5\mu^{1\over4}\simeq0.82$ for
$\mu=5M_J/1.6M_\odot=0.003$.  Thus, our putative four planet
configuration is likely just marginally stable for the age of the
system.  We then ask if this orbital configuration may account for the
entire 8--80 AU gap.  By construction, planet 1 and planet b shepherd
the edges of the warm and cold belts, respectively.  We calculate
that, for circular planetary orbits, the exterior chaotic zone of
planet 1 extends to $\sim$13 AU, the interior chaotic zone of planet 2
extends to $\sim$14 AU, the exterior chaotic zone of planet 2 extends
to $\sim$24 AU, the interior chaotic zone of planet 3 extends to
$\sim$26 AU, the exterior chaotic zone of planet 3 extends to $\sim$44
AU.  Thus, within the observational and theoretical uncertainties, we
consider that this configuration of four equal-mass planets, each of
mass 5 $M_J$, in nearly circular orbits of semi-major axes 11 AU, 19
AU, 34 AU and 62 AU, can account for the entire gap between the warm
and cold belts, and is also just marginally stable for the age of the
system.  This four-planet configuration is illustrated in Figure 9 as
``system D''.  We note that a four-planet system is unlikely to be
stable if the planet masses exceed $\sim$5 $M_J$. We also note that HR
8799 planets are not stable under these stability criteria;
\citet{gozdziewski14} propose a multiply resonant orbital
configuration to ensure dynamical stability of this system. Similarly,
higher mass, coplanar planetary configurations of four planets in HD
95086 could be stable if the orbital parameters were fine-tuned to
involve mean motion resonances.

For the directly-imaged four-planet system of HR 8799, a massive cold
debris disk exceeding 10\% of the outermost planet's mass has been
suggested to stabilize otherwise unstable configurations for several
Myr \citep{moore13}.  In the case of HD 95086, the mass of the cold
debris belt in particles up to 1 mm size is $\sim$0.2 $M_\oplus$;
extrapolating to km-size bodies using a $-$3.5 size power-law yields a
total cold disk mass of $\sim$200 $M_\oplus$.  If this debris disk
mass estimate is robust, the cold disk might play a role in
stabilizing a dynamically packed planetary system with multiple
massive planets similar to that of HR 8799.

\section{Conclusions}

We present a detailed study of the debris disk around HD 95086, a
young star that also harbors a directly imaged planet. We find: 

1.) The observed disk SED of HD 95086 is best
described with two dust temperatures: a warm component at $\sim$175 K
and a cold component at $\sim$55 K. There is a possible hotter, but
much fainter component at $\sim$300 K, suggesting the presence of
debris in the terrestrial planet zone that needs future confirmation.
 
2.) We detect an emission feature in the low-resolution MIPS-SED
data. The feature peaks at 69.5$\pm$0.5 \mm, and is not spectrally
resolved. Its integrated luminosity and peak wavelength are consistent
with a mineralogical feature originating from iron-poor, crystalline
olivine that has been detected in many protoplanetary disks and in the
$\beta$ Pic debris disk. HD 95086 is the second debris disk, after
$\beta$ Pic, to show a feature. We estimate the mass of crystalline
grains to be $\sim$6.2$\times$10$^{25}$ g, accounting for $\sim$5\%
abundance in the observed dust debris. Because of the high temperature
of crystallization, the substantial mass required to produce this
feature must have been transported from regions close to the star, or
heated in the core of a planetary body that have been disrupted. The
latter is consistent with the fact that this feature only detected in
dustiest systems in young debris disks where collisional rates are
high.

3.) Detailed analysis of the far-infrared resolved images suggests
the debris emission can be traced up to radii of $\sim$9\arcsec\ (800
AU) where $\sim$half of the emission is within the beam size (radii
$\lesssim$250 AU) and not resolved. Our SED models suggest that the
cold component has an outer boundary (radius) less than 200 AU,
contradicting our imaging analysis. To reconcile the SED models and
the extended emission, a third component in the SED models is
needed. The extended third component is likely to be a disk halo,
similar to the ones that have been found in Vega and HR 8799
\citep{su05,su09,matthews14}, made of grains closer to or smaller than
the blowout size that stellar radiation pressure places them in highly
eccentric or hyperbolic orbits. We confirm previous determinations
that the resolved structure is at an inclination of
25\arcdeg$\pm$5\arcdeg\ and a position angle of 115\arcdeg$\pm$10\arcdeg,
and find that the eccentricity of the disk halo is small, $e<$0.3.

4.) We compare the derived properties of the three components in HD
95086 to those of HR 8799 \citep{su09}, and find a striking similarity
in debris structures. Both systems possess warm and cold excesses with
an orbital ratio of $\sim$10, similar to the ratios between Asteroid
and Kuiper belts in our Solar System and some extrasolar debris disks
\citep{su13,kennedy14}. Most importantly, both HD 95086 and HR 8799
systems also possess a bright disk halo, suggesting an elevated level
of activities in the leftover planetesimals. It is interesting to note
that the difference in dust level (fractional luminosity) is $\sim$10
times in both warm and cold excesses between HD 95086 and HR 8799
where HD 95086 is dustier.

5.) We explore the possible planetary configurations present in the HD
95086 system using dynamical constraints from the debris distribution
and currently known 5 $M_J$ planet at a astro-centric distance of
61.5 AU in coplanar cases. For the case of a single planet system, the
eccentricity of HD 95086b is required to be $\approx$0.7 to maintain
the large gap. Although we cannot rule out the single planet case, the
origin of its high eccentricity requires an additional
explanation. For the case of a two-planet system, the current mass
limit (21$M_J$ at a astro-centric distance of $\sim$16 AU) rules out
two planets on low eccentricity ($e<$0.1) orbits; however, two 5 $M_J$
planets on modest ($e\sim$0.3) eccentricity orbits are marginally
stable for $\sim$20 Myr and can maintain the large gap.

6.) We further estimate the stability of packed configurations for
three and four equal-mass planets on low eccentricity ($e<$0.1)
orbits. In the case of three planets, the planets need to be as
massive as 7 $M_J$ (the maximum mass for HD 95086b) and at semi-major
axes of $\sim$12 AU, 26 AU, and 56 AU to be dynamically stable and to
maintain the gap. Three planets of much smaller mass would be less
likely to maintain the large gap.  We also show that the HD 95086
system could host four low-eccentricity, massive planets between its
warm and cold belts, similar to the iconic HR 8799 system; however,
barring fine-tuning with mean motion resonances, these four planets
would have masses $\lesssim$5 $M_J$ in order to be stable over the
system's age ($\sim$20 Myr).

Finally, our results also illustrate resolved imaging (even marginally
resolved in this case) can provide much more insight if accompanied by
proper SED modeling.  Since both HD 95086 and HR 8799 systems are in
the epoch when active dynamical processes like final settling of giant
planet migration and terrestrial planet formation are in play, the
observed difference in the dust level probably reflects the age
difference, i.e., HD 95086 is a younger analog of HR 8799, rather than
initial conditions. Future higher sensitivity data in the properties
of the debris components will provide a better time resolution in this
active planet formation epoch.  Among the currently known 9 directly
imaged extrasolar planets\footnote{HR 8799 bcde: \citet{marois10},
Fomalhaut b: \citet{kalas08}, $\beta$ Pic: \citet{lagrange09}, HD
95086 b: \citet{rameau13a}, GJ 504 b: \citet{kuzuhara13}, HD 106906 b:
\citet{bailey14}.} (masses less than 13 $M_J$), the host stars also
possess bright debris disks except for GJ 504, implying a strong
connection between bright debris disks and directly imaged
planets. The demography of planetary systems revealed by indirect
methods (radial velocity, transit, micro-lensing and presence of
debris) suggests the formation and evolution of planetary systems have
multiple paths. Therefore, finding common features among planetary
systems is the first step to better understand their formation and
evolution.

\acknowledgments

We thank the anonymous referee for his/her rapid report and
suggestions led to a better presentation of the paper. We thank B. de
Vries for providing the PACS spectrum of $\beta$ Pic. KYLS
acknowledges the support provided by the NASA Astrophysics Data
Analysis Program through grant \# NNX11AF73G. SJM acknowledges funding
from NASA grants \#NNX13AO65H and \#NNX14AG93G.  RM acknowledges
funding from NASA grant \#NNX14AG93G and NSF grant \#AST-1312498. ZB
is funded by the Deutsches Zentrum f\"ur Lufund Raumfahrt (DLR).

\end{document}